# Momentum transfer by an internal source of ionizing radiation — an important feedback process during galaxy formation?!

Martin G. Haehnelt

*Max-Planck-Institut für Astrophysik, Karl-Schwarzschild-Straße 1, 85748 Garching b. München, Germany, e-mail: haehnelt@mpa-garching.mpg.de*

## ABSTRACT

*The role of momentum transfer ("radiation pressure") due to an internal source of ionizing radiation for the formation of baryonic structures is investigated. Fully-ionized self-gravitating gaseous objects can be radiation-pressure supported on a characteristic length scale $D_{\rm rp} \sim 100\,{\rm pc} - 3\,{\rm kpc}$. On smaller scales momentum transfer due to ionizing radiation will be the dominant force and for spherical collapsing objects of mass $\lesssim 10^{10}\, M_\odot$ a complete bounce is possible. A population of massive stars and/or accretion onto a central compact object are natural sources of ionizing radiation in newly-forming baryonic structures. Radiation pressure is therefore likely to play an important role for the dynamical and thermal evolution of the intergalactic medium and the mass-to-light ratio of small galaxies. The effect will be especially important for hierarchical cosmogonies where galactic structures build up by merging of smaller objects. Radiation pressure due to ionizing radiation might furthermore be responsible for substructures of size $D_{\rm rp}$ in the stellar component of large galaxies.*

## 1 INTRODUCTION

In the canonical (gravitational-instability) picture for structure formation (baryonic) structures form via the growth and collapse of small density inhomogeneities in an initially rather homogeneous universe. It is well known that then for gas with densities and (virial) temperature typical for objects with masses of $\sim 10^7 - 10^{13}\, M_\odot$ virializing at moderate redshifts bremsstrahlung and line-cooling are the dominant cooling processes and that the cooling time is shorter than the free-fall time (Binney 1977, Rees & Ostriker 1977, Silk 1977). The exact mass limits depend of course on virialization redshift, gas density of the universe, metal content and density distribution within the object. Rees & Ostriker also showed that for contracting objects the maximum possible momentum input of radiation due to the release of gravitational energy is in any case insufficient to balance the gravitational attraction. No pressure support is therefore possible in such objects without additional input of heat or momentum. Natural sources for such an energy/momentum input are massive stars and/or accretion onto a central compact object. The importance of such "feedback" processes for the cooling and star formation history of the intergalactic medium (IGM) and the structure and formation of galaxies is more and more realized (White & Frenk 1991, Katz 1992, Cen 1992, Cen & Ostriker 1993, Navarro & White 1993, Metzler & Evrard 1994, Mihos & Hernquist 1994, Steinmetz & Müller 1994). Most widely studied are the effects of supernova ( *e.g.* Dekel & Silk 1986).

In this paper I will investigate another feedback effect of massive stars (or accretion onto a central compact object) which so far has got little or no attention, the momentum transfer to the gas due to ionizing radiation. It will turn out that this momentum transfer can be astonishingly efficient and that global pressure support and even a radiation-pressure driven bounce are possible in sub-galactic structures. In section 2 I first discuss the simple case of a homogeneous self-gravitating sphere of hydrogen of uniform density with a central internal source of ionizing radiation. This serves to explain the basic mechanism and to derive a fundamental characteristic length scale for pressure support by ionizing radiation. Then I investigate the dynamical evolution of a fully-ionized object. Later on more complicated distributions of density and emissivity, the role of resonant line scattering and the special case of a disc-like geometry are briefly studied. In section 3 I investigate under which physical conditions the effect is important and which role it plays for the formation of baryonic structures. Some general cosmological implications are also discussed. Section 4 contains the conclusions.

## 2 MOMENTUM-TRANSFER BY AN INTERNAL SOURCE OF IONIZING RADIATION

### 2.1 The basic mechanism — a point source at the centre of a homogeneous sphere of hydrogen with uniform density

Momentum transfer by radiation is one of the basic astrophysical processes. Most widely known is probably the concept of the Eddington luminosity, which for an object of mass $M$ can be written as

$$L_{\text{edd}} = \frac{4\pi \, G \, M \, m_p \, c}{\sigma_T}, \tag{1}$$

where $m_p$ is the proton mass and $\sigma_T$ the cross section for Thomson scattering. The Eddington luminosity is the critical luminosity at which the flux of outward momentum transferred to the gas by Thomson scattering balances the attracting gravitational force of a self-gravitating object. However, the Salpeter timescale, the characteristic lifetime of an object radiating at the Eddington limit and transforming its rest-mass energy into radiation with efficiency $\epsilon_{\text{rad}}$ is only

$$t_{\text{Salp}} = \frac{\epsilon_{\text{rad}} \, M \, c^2}{L_{\text{edd}}} \sim 4.5 \times 10^8 \epsilon_{\text{rad}} \text{ yr} \tag{2}$$

This is much shorter than typical cosmological timescales and even shorter than typical dynamical timescales of objects virializing at moderate redshifts unless $\epsilon_{\text{rad}}$ is unreasonably high. Momentum transfer by Thomson scattering of radiation produced by feedback processes due to massive stars is therefore generally not efficient enough for the corresponding radiation pressure to be dynamically important during galaxy formation. However, the fact that absorption cross-sections for ionizing radiation are typically a factor $\sim 10^7$ larger than the cross section for Thomson scattering suggests that ionizing radiation is a much more efficient means to transfer momentum to the gas and might have important dynamical effects on the gas.

In this section I will use a simple example — a homogeneous self-gravitating sphere of hydrogen with uniform mean density containing a central point source of ionizing radiation — to demonstrate under which conditions momentum transfer by absorption of ionizing radiation is sufficient for pressure support. If collisional ionization is neglected the recombination rate of ionized hydrogen sets a basic upper limit to the rate at which momentum can be transferred to the gas by ionizing photons. In the case of a ionizing point source at the centre of a sphere the exerted pressure is maximally anisotropic and the resulting time-averaged radially outward directed force on a hydrogen atom is given by

$$F_{\text{ion}}^{\text{max}}(r) = \frac{[x(r)]^2 \, \alpha_{\text{rec}} \, h\nu_{\text{ion}} \, n_H}{c}, \tag{3}$$

where $x(r) = n_{HII}/n_H$ is the ionized fraction, $\alpha_{\text{rec}}$ is the effective hydrogen recombination coefficient and $\nu_{\text{ion}}$ is the frequency of a typical ionizing photon. The gravitational force on the hydrogen atom is given by

$$F_{\text{grav}}(r) = \frac{G\,M(r)\,m_p}{r^2} = \frac{4\pi\,G\,m_p^2\,n_H\,r}{3}, \tag{4}$$

where $M(r)$ is the integrated mass within a distance $r$ to the centre of the sphere and the second relation holds for uniform mean density. The ionized fraction will generally be decreasing with increasing radius and comparison of equation (3) and (4) shows that there is a maximum distance for which pressure support is possible which is given by the implicit equation

$$R_{\text{rp}}^{\text{hom}} = \frac{3\,[x(R_{\text{rp}}^{\text{hom}})]^2\,\alpha_{\text{rec}}\,h\nu_{\text{ion}}}{4\pi\,c\,G\,m_p^2}. \tag{5}$$

In the following I will be mainly interested in the case in which the gas is highly ionized everywhere and $x \approx 1$ throughout the object will be assumed for the rest of the paper. The maximum distance of equation (5) correspond then to a characteristic length scale

$$D_{\text{rp}} = \frac{3\,\alpha_{\text{tot}}\,h\nu_{\text{ion}}}{4\pi\,c\,G\,m_p^2} \sim 100 - 150\,\text{pc} \tag{6}$$

The gas is here assumed to be optically thin to ionizing radiation ($\alpha_{\text{rec}} = \alpha_{\text{tot}}$, where $\alpha_{\text{tot}}$ is the total hydrogen recombination rate). Furthermore $10^4$ K and $1 - 1.5$ Rydberg are assumed as the typical temperature for a gas heated by photoionization and the typical energy of an ionizing photon. Note that for a homogeneous uniform gas distribution $D_{\text{rp}}$ only depends on fundamental atomic physics constants.

## 2.2 Stability, disruption and bounce of collapsing objects

Equations (3) and (4) demonstrate a dynamically important property of radiation pressure exerted by a source of ionizing radiation at the centre of a self-gravitating object. Independent of the radial distribution of the mass the transferred momentum flux increases faster for smaller radii than the gravitational force ($F_{\rm ion}(r)/F_{\rm grav}(r) \sim (R_{\rm rp}/r)$). Let us now consider what happens if a source of ionizing radiation is turned on in an object of radius $R_{\rm ion}$. For objects with $R_{\rm ion} \sim R_{\rm rp}$ a quasi-equilibrium state is possible which can last as long as the luminosity of ionizing radiation is sufficient for full ionization ($L_{\rm ion} \geq L_{\rm ion}^{\rm min}$). However, equilibrium can only be attained at the outer edge and a significant redistribution of matter will take place within the object. For larger objects no global pressure support is possible and for smaller objects the flux of momentum due to ionizing radiation significantly exceeds the gravitational force.

In the following I will discuss the latter case in somewhat more detail (see section 2.4 for some further discussion of larger objects). An object with $R_{\rm ion} < R_{\rm rp}$ will first become spherical and depending on the radial distribution and emissivity of ionizing radiation material might be redistributed toward the outer edge (see also section 2.3.1). Then a global radiation-pressure-driven expansion will start. For the outermost mass shell of radius $R$ equation (3) and (4) give the following equation of motion

$$\ddot{R} = \frac{GM}{R_{\rm rp}^2} \left(\frac{R_{\rm rp}}{R}\right)^3 \left(1 - \frac{R}{R_{\rm rp}}\right), \tag{7}$$

where $M$ is the total enclosed mass. This can be integrated to obtain the expansion velocity

$$\dot{R} = \left(\frac{GM}{R_{\rm rp}} \left[\left(\frac{R_{\rm rp}}{R}\right)^2 \left(\frac{R}{R_{\rm rp}} - \frac{1}{2}\right) + \left(\frac{R_{\rm rp}}{R_{\rm ion}}\right)^2 \left(\frac{1}{2} - \frac{R_{\rm ion}}{R_{\rm rp}}\right)\right]\right)^{1/2}, \tag{8}$$

where zero velocity is assumed at the onset of ionization ($\dot{R}(R_{\rm ion}) = 0$). An investigation of equation (8) for large R shows that the object will be completely disrupted ($\dot{R}(R = \infty) = v_\infty \geq 0$) if $R_{\rm ion}/R_{\rm rp} \leq 1/2$. Similarly, for an object which is in free-fall collapse when ionization sets in ($\dot{R}(R_{\rm ion}) \sim -(GM/R_{\rm ion})^{1/2}$) a bounce and subsequent disruption will occur if $R_{\rm ion}/R_{\rm rp} \leq 1/4$.

This behaviour is remarkably different from the usual situation in stellar-size radiation-pressure-dominated self-gravitating objects which are normally collisionally ionized and dominated by Thomson scattering. These have properties close to that of a polytrope with polytropic index 4/3 and are marginally stable without an equilibrium length scale.

## 2.3 Complicating effects in a more realistic situation

*2.3.1 Extended source of ionizing radiation*

In section 2.1 the radiation pressure was estimated for the rather simple case of a central point source (which might resemble a central accreting object). As mentioned above in this case the pressure is maximally anisotropic and has only a radially outward component. However, if the ionizing radiation is mainly produced by massive stars, an extended source of ionizing radiation should be assumed and the situation becomes more complicated. The hydrogen atom will now absorb ionizing photons from different directions. For a spherical configuration the resulting force will still point radially outward, but it will be reduced compared to the maximum force given by equation (3) and the pressure will be less anisotropic. If the optical depth for ionizing radiation is small throughout the object, the probability for an ionizing photon to originate from a region $dV(\mathbf{r})$ of the object is $\propto j_{\mathrm{ion}}(\mathbf{r})\,dV(\mathbf{r})$, where $j_{\mathrm{ion}}(\mathbf{r})$ is the emissivity of ionizing radiation at point $\mathbf{r}$. A position-dependent reduction factor can then be defined for an extended source of radiation, which is given by

$$F_{\mathrm{ion}}(r) = A_{\mathrm{ext}}(r)\, F_{\mathrm{ion}}^{\max},$$
$$A_{\mathrm{ext}}(r) = \int_{\mathrm{sphere}} j_{\mathrm{ion}}(\mathbf{r'}) \frac{(\mathbf{r'}-\mathbf{r})\cdot\mathbf{r}}{|\mathbf{r'}-\mathbf{r}|\,|\mathbf{r}|}\,dV(\mathbf{r'}) \Big/ \int_{\mathrm{sphere}} j_{\mathrm{ion}}(\mathbf{r'})\,dV(\mathbf{r'}), \qquad (9)$$

where the coordinate system was chosen so that its centre coincides with the centre of the sphere. It is now illustrative to investigate the special case of spatially constant emissivity, which can be solved analytically

$$A_{\mathrm{ext}}(r) = \frac{r}{R}\left(1 - \frac{1}{5}\left(\frac{r}{R}\right)^2\right), \qquad (10)$$

where $R$ is again the radius of the outermost mass shell. The force due to ionizing radiation will be reduced by a small factor at the outer edge of the sphere and it will decrease approximately linearly towards the centre ($F_{\mathrm{ion}}(r) \propto r/R$). Comparing equation (3), (4) and (5) shows that then $F_{\mathrm{ion}}(r)/F_{\mathrm{grav}}(r) \sim R/R_{\mathrm{rp}} = \mathrm{const.}$. In this case there exists an equilibrium solution for objects with radius $R \sim R_{\mathrm{rp}}$, where the gravitating force balances the flux of transferred momentum throughout the object and no redistribution of matter will take place. However, in a realistic situation the gas cloud should be somewhat centrally condensed and as $j_{\mathrm{ion}}$ is likely to be correlated with the gas density (probably even in a non-linear fashion) it will also be centrally condensed. This will enhance $F_{\mathrm{ion}}$ at small radii and for objects with radii of order $R_{\rm rp}$ a self-regulating mechanism might be established which inhibits the formation of a strong central condensation.

*2.3.2 Centrally condensed and inhomogeneous mass distributions*

Equation (3) and (4) show that at a given radius the ratio $F_{\rm ion}/F_{\rm grav} \propto (3-\gamma)$ for a density distribution $n_{\rm H} \propto r^{-\gamma}$ and $R_{\rm rp}$ is *e.g.* reduced by a factor of 3 for an isothermal sphere compared to a sphere of uniform density. Furthermore, as discussed above, the radial distribution of mass and ionizing emissivity might be correlated. While the resulting changes are rather small in both of these cases, an inhomogeneous gas distribution will have a much stronger effect. As discussed in section 2.1, the average force exerted on an individual proton is determined by the recombination rate if collisional ionization is neglected. For an inhomogeneous object the ionizing force will therefore be enhanced by a clumping factor

$$F_{\rm ion}^{\rm inhom} \sim \delta_\rho F_{\rm ion}^{\rm hom}, \qquad \delta_\rho = \langle n^2 \rangle / \langle n \rangle^2. \qquad (11)$$

and $R_{\rm rp}^{\rm inhom} \sim \delta_\rho R_{\rm rp}^{\rm hom}$. For a two-phase medium with a tenuous hot and a dense cool phase in pressure equilibrium this clumping factor will be given by $\delta_\rho \sim f_{\rm hot} T_{\rm hot}/T_{\rm cool}$, where $f_{\rm hot}$ and $T_{\rm hot}$ are mass fraction and temperature of the hot phase and $T_{\rm cool}$ is the temperature of the cool phase. Reasonable values of the clumping factor should lie in the range $1 \lesssim \delta_\rho \lesssim 30$ and the typical length scale for radiation pressure due to ionizing radiation will be significantly enhanced to $D_{\rm rp} \sim 100\,{\rm pc} - 3\,{\rm kpc}$.

*2.3.3 (Ly $\alpha$-) Scattering and momentum transfer due to other elements*

So far ionization of hydrogen was considered as the only means of momentum transfer by radiation. However, resonant line scattering is likely to provide additional radiation pressure. In a realistic situation the gas will be a mixture of hydrogen, helium and metals providing a complicated structure of atomic levels. A detailed treatment of the radiative transfer problem is difficult and I will only make some rather short comments on the most important "secondary" effects.

Recombination of hydrogen to levels above the ground level (with recombination rate $\alpha_2$) will lead to production of Ly-$\alpha$ photons and there will be an additional isotropic component of radiation pressure $P_{{\rm Ly}-\alpha} \sim U_{\rm trap}/3$, where $U_{\rm trap}$ is the energy density of trapped Ly-$\alpha$ photons. There will still be a radially outward directed resulting force

$$F_{\text{Ly}-\alpha} = -\frac{1}{n_{\text{H}}} \nabla_r (P_{\text{Ly}-\alpha}) \sim -\frac{1}{n_{\text{H}}} \nabla_r \left( \frac{1}{3} \alpha_2 \, n_{\text{H}}^2 \, h\nu_{\text{Ly}-\alpha} \, \frac{r}{c} \, \frac{t_{\text{trap}}}{t_{\text{light}}} \right)$$
$$\sim \frac{2\gamma - 1}{3} \frac{\alpha_2}{\alpha_{\text{tot}}} \frac{h\nu_{\text{Ly}-\alpha}}{h\nu_{\text{ion}}} \frac{t_{\text{trap}}}{t_{\text{light}}} F_{\text{ion}}^{\text{max}}, \tag{12}$$

if the radial density profile is centrally concentrated (Elitzur & Ferland 1986, Braun & Dekel 1990). This force will be larger than $F_{\text{ion}}^{\text{max}}$ by a factor $A_{\text{scatt}} \sim (\gamma/6)(t_{\text{trap}}/t_{\text{light}})$, where $t_{\text{trap}}$ is the typical trapping time of an Ly-$\alpha$ photon, $t_{\text{light}} = r/c$ is the light travel time for direct escape from the centre of the object and $(\alpha_2/\alpha_{\text{tot}})(h\nu_{\text{Ly}-\alpha}/h\nu_{\text{ion}}) \sim 1/4$ is assumed. For a fully ionized object the typical optical depth for Ly-$\alpha$ scattering at the line centre is $\sim 10^3 - 10^4 (L_{\text{ion}}/L_{\text{ion}}^{\text{min}})^{-1}$. However, even for such considerable optical depths it is rather difficult to trap Ly-$\alpha$ radiation efficiently. This is due to the fact that photons are quickly Doppler-shifted into the wings of the line, where the absorption cross section drops dramatically. Thermal motions at $10^4\,\text{K}$ already limit the trapping time to $t_{\text{trap}}/t_{\text{light}} \sim 10 - 15$, only weakly dependent on $\tau_{\text{Ly}-\alpha}$ (e.g. Bithell 1990) and Ly-$\alpha$ scattering will at most enhance the effective radiation pressure by a factor $\sim 3 - 5$. Random motions within an object further reduce this value and trapping of Ly-$\alpha$ photons is therefore unlikely to be very important in objects with virial velocities significantly in excess of $10\,\text{km s}^{-1}$.

Momentum transfer due to ionization of other elements than hydrogen will not be relevant as the comparatively small abundance cannot be compensated by recombination rates which are only for a few species slightly larger than that of hydrogen. However, line scattering by metals might be important. Its role will crucially depend on the metallicity, on the detailed ionization structure of the gas and therefore also on the spectral distribution of the radiation emitted by the internal source.

### 2.4 Disc-like geometry

Even though radiation pressure due to ionizing radiation can only be dynamically dominant in self-gravitating objects with radius $R_{\text{ion}} \lesssim R_{\text{rp}}$, it might still play an interesting role in angular-momentum supported objects with larger radii which have collapsed into a disc. The average vertically outward directed force onto a proton exerted by ionizing radiation is then given by

$$F_{\text{ion}}^{\text{disc}}(z) \sim \frac{\delta_\rho \, \alpha_{\text{rec}} \, h\nu_{\text{ion}} \, n_H}{c} \frac{2z}{D_{\text{ion}}}, \tag{13}$$

where z is the distance from the central plane and $D_{\text{ion}}$ is the typical mean free path of an ionizing photon within the disc. Note the additional factor $2z/D_{\text{ion}}$ compared to equation

(3) which is due to the necessarily <u>extended</u> source of ionizing radiation. The vertical gravitating force for an infinitely extended disc is given by

$$F_{\rm grav}^{\rm disc}(z) = 2\pi G\,\Sigma(z) = 4\pi\,G\,m_p^2\,n_H\,z, \tag{14}$$

where $\Sigma(z) = \int_0^z 2\,n_H\,{\rm d}z$ is the vertically integrated surface mass density and the second relation holds again for uniform mean density. Comparison of equation (13) and (14) shows that $F_{\rm ion}$ and $F_{\rm grav}$ would have the same scaling with disc height $H$ and no equilibrium were possible if $D_{\rm ion}$ were independent of $H$. However, $D_{\rm ion} \sim H^2/H_{\rm ion}^{\rm min}$, where $H_{\rm ion}^{\rm min} \sim 15\,\delta_\rho\,(\Sigma(H)/\,1\,{\rm g\,cm^{-2}})(\mu_{\rm ion}/\mu_{\rm edd})^{-1}$ pc is the minimum disc height for which the disc can be kept fully ionized with a given surface brightness of ionizing radiation $\mu_{\rm ion}$. There is therefore a maximum disc height, $H_{\rm rp} \sim 2\,R_{\rm rp}/3$, for which momentum transfer due to ionizing radiation can balance gravity and where $\mu_{\rm ion} \sim \mu_{\rm ion}^{\rm min}(H_{\rm rp})$. For a given $\mu_{\rm ion} \gtrsim \mu_{\rm ion}^{\rm min}(H_{\rm rp})$ equilibrium is attained at a disc height

$$H_{\rm eq} \sim (H_{\rm rp}\,H_{\rm ion}^{\rm min})^{1/2} \sim 300 \left(\frac{\delta_\rho}{10}\right)\left(\frac{\Sigma(H)}{1\,{\rm g\,cm^{-2}}}\right)^{1/2}\left(\frac{\mu_{\rm ion}}{\mu_{\rm edd}}\right)^{-1/2}\,{\rm pc}. \tag{15}$$

Note that $H_{\rm eq}$ decreases with increasing $\mu_{\rm ion}$.

# 3 RADIATION PRESSURE AND GALAXY/STRUCTURE FORMATION

## 3.1 The "regime" of momentum transfer by ionizing radiation

I will now discuss under which conditions radiation pressure due to ionizing radiation might be relevant. For this purpose it is useful to use the mean density and the radius to characterize an object. Figure 1 illustrates the basic relations assuming a self-gravitating inhomogeneous sphere of hydrogen ($\delta_\rho = 10$) with uniform mean density which is fully ionized with the minimum required luminosity ($L_{\rm ion} \sim L_{\rm ion}^{\rm min}$). The un-shaded region on the left hand side indicates the range of densities and radii where radiation pressure will be dynamically dominant. The region is bounded by the characteristic radius beyond which the gravitational attraction cannot be balanced by momentum transfer due to ionizing radiation (thick solid line), a maximum density, $\sim 10^4\,\delta_\rho^{-1}\,{\rm cm^{-3}}$, above which the minimum luminosity necessary to ionize the object exceeds the classic Eddington luminosity (thick short-dashed line) and the minimum total column density, $\sim 10^{17}$ cm$^{-2}$, below which the optical depth for ionizing radiation drops below unity even if the object is completely neutral (lower thick long-dashed line). While the first two of these lines are strict limits, radiation pressure support due to ionizing radiation is also possible below the line where $y \sim 1/2$, but the ionizing luminosity has then in any case to be (much) larger than $L_{\rm ion}^{\rm min}$. There is an upper mass limit of $\sim 10^{10}$ $M_\odot$ for objects falling into the un-shaded region. The neutral fraction depends on the total column density and the luminosity of ionizing radiation, $y = n_{HI}/n_H \sim (N_H/10^{17}\,{\rm cm}^{-2})^{-1}\,(L^{\rm ion}/L_{\rm ion}^{\rm min})^{-1}$. For relevant column densities this is much smaller than unity and the assumption of high ionization $x = 1 - y \approx 1$ is generally well justified. The necessary ionizing luminosity is given by $L_{\rm ion}^{\rm min} \sim \delta_\rho\,(n_H/10^4\,{\rm cm}^{-3})L_{\rm edd}$.

As mentioned above, momentum transfer due to ionizing radiation could also be relevant for the way the gas collapses into a disc in more massive angular-momentum supported objects. For such disc-like objects total surface mass density and disc height are the appropriate parameter and Fig. 2 shows the relevant conditions in a similar fashion as Fig. 1. The un-shaded region on the left hand side indicates the range of surface mass densities and disc heights where radiation pressure will be dynamically important in a self-gravitating disc fully ionized with $\mu_{\rm ion} \sim \mu_{\rm ion}^{\rm min}$. The region is bounded by the maximum height $H_{\rm rp}$, a maximum density for which the surface brightness necessary to ionize disc is below the Eddington surface brightness and the minimum column density. Dashed-dotted lines show the equilibrium disc height as given in equation (15).

Momentum transfer due to ionizing radiation will be relevant for the formation of baryonic structures mainly for two reason:

- The baryons in collapsing objects of baryonic masses up to $\lesssim 10^9$–$10^{10}$ $M_\odot$ will bounce if efficient formation of massive stars or formation of a supermassive central compact object leads to full ionization at a radius smaller than $D_{\rm rp}$. As will be discussed in section 3.2., this might be a common process in the large number of objects of this mass range which are expected to form at moderate redshifts in a hierarchical cosmogony.

- During the formation of galaxies substructure with scale $\sim D_{\rm rp}$ might attain equilibrium for a few dynamical times. This might be sufficient for the formation of a distinct stellar subsystem and might *e.g.* be relevant for the formation of small bulges in late type spiral galaxies and the thick disc of our galaxy (see section 3.3. for a further discussion).

## 3.2 Suppression of the collapse of systems with scale $\lesssim \mathbf{D_{rp}}$ and its implications for hierarchical cosmogonies

It is widely assumed that the universe is dynamically dominated by collisionless dark matter. Most currently popular cosmogonies furthermore invoke a hierarchical clustering process for the build-up of structures; larger structures form by merging of smaller systems. In such models there is a typical mass scale $M_*(z)$ and most of the mass is contained in objects in a rather small mass range around $M_*$ spanning a few decades in mass. The exact normalization and the strength of the evolution depend of course on the special cosmological model chosen. For currently popular cosmogonies $M_*(z = z_8) \sim 10^8 \, M_\odot$ at a redshift $3 \lesssim z_8 \lesssim 20$. The formation of objects around $10^8 \, M_\odot$ should therefore be common at moderate redshift and as I will show below a radiation-pressure driven bounce is likely to occur in these objects. To see this let us consider the canonical picture for the evolution of an individual density inhomogeneity which decouples from the Hubble-flow. While the collisionless dark matter component virializes at half the turn-around radius and forms a (probably isothermal) dark-matter halo, the baryons are able to cool and will contract further until angular-momentum support sets in. Star formation will start when the baryons become self-gravitating. In an isothermal halo this occurs after collapse by a factor $f_{\rm bar}^{-1}$ in radius at a typical density

$$n_{\rm sg} \sim 2 \times 10^{-1} \left(\frac{f_{\rm bar}}{0.05}\right)^{-2} (1+z)^3 \, {\rm cm}^{-3}, \qquad (16)$$

where $f_{\rm bar}$ is the baryonic mass fraction. This density is indicated in Figure 1 for different redshifts (dotted lines). For a given baryonic mass this corresponds to a typical radius

$$R_{\rm sg} \sim 1.5 \left(\frac{M_{\rm bar}}{10^8 \, M_\odot}\right)^{1/3} \left(\frac{f_{\rm bar}}{0.05}\right)^{2/3} (1+z)^{-1} \, {\rm kpc} \qquad (17)$$

Self-gravity and therefore a possible ionization of the baryonic concentration at the centre of the halo occurs just at those radii where a radiation-pressure driven bounce is possible and the collapse could be reverted. There are now two questions to be answered:

- Is such a bounce energetically possible?

- Is it possible to form enough massive stars for full ionization?

It turns out that both questions contain no serious constraint for small dark-matter-haloes. Approximately a fraction $(R_{\rm ion}/R_{\rm rp})^{1/2}(v_\infty/c)(L_{\rm ion}/L_{\rm ion}^{\rm min})^{-1}$ of the energy in ionizing photons is transferred into kinetic energy of the gas, where $v_\infty \sim (R_{\rm rp}/R_{\rm ion})^{1/2} \, v_{\rm vir}$ is the final velocity of the gas accelerated by ionizing radiation (equation (8)), $v_{\rm vir} \sim (GM/R_{\rm ion})^{1/2}$ is the virial velocity of the dark-matter halo and ionization is assumed to occur on a dynamical timescale when the gas becomes self-gravitating). A bounce can occur if the fraction of the rest mass converted into ionizing radiation is $\gtrsim (v_{\rm vir}/c)(L_{\rm ion}/L_{\rm ion}^{\rm min})$. This is only a moderate constraint for small dark-matter haloes. To answer the second question a look at Fig. 1 is instructive. Typically $L_{\rm ion}/L_{\rm edd} \sim 10^{-2}$ is sufficient for full ionization. A simple estimate assuming that all ionizing photons are produced in stars more massive than $10\,M_\odot$ with an emission rate $Q_{\rm ion} \sim 8 \times 10^{47} (M/10\,M_\odot)^2\,{\rm s}^{-1}$ (Bithell 1990) gives $L_{\rm ion}/L_{\rm edd} \sim 5 \times 10^{-3}\,f_*$ for an coeval burst of star formation with an Salpeter initial mass function (IMF), mass limits $0.1\,M_\odot \leq M \leq 100\,M_\odot$ and a star formation efficiency $f_*$. For an IMF biased towards massive stars a factor $10-100$ higher value would be possible. Even a rather low star formation efficiency would then be sufficient for ionization and a major fraction of the self-gravitating gas at the centre can be blown off in a starburst with an IMF which is only moderately biased towards massive stars. Furthermore, the expanding gas sphere will sweep up the surrounding material like a supermassive expanding supernovae remnant and an even larger amount of gas contained in the outer parts of the halo can be removed in a momentum-conserving expansion phase.

What is now the main difference to what is generally considered as the most important feedback process, heating by expanding supernova remnants? It is the high efficiency with which momentum can be transferred to <u>all</u> the remaining gas. Energetically the bilance is very similar, but in the case of supernovae heating a considerable fraction of all gas will be contained in the cool dense shells of the remnants, which loose their bulk velocity by collision with neighbouring remnants. This cool dense gas is very hard to remove from the halo by subsequent supernovae and while a considerable fraction of the gas is blown off as a hot wind the major fraction of the gas can probably not be removed from a dark-matter halo in this way. The global heating efficiencies which are often used to model the feedback effect of supernovae in numerical simulations or analytical calculations might therefore strongly overestimate their effect. However, a radiation-pressure driven bounce can completely remove the baryons from haloes with baryonic masses up to $10^9\,M_\odot$ and under special conditions maybe even $10^{10}\,M_\odot$. Such a mechanism has similar cosmological implications as those generally assumed for supernovae. It will significantly contribute to the heating of the IGM and it might be an alternative explanation for the strong rise of the mass-to-light ratio of galaxies towards low luminosities and the related problem why the galaxy luminosity function is considerable flatter at the faint end than expected from simple hierarchical clustering models ( *e.g.* White & Frenk 1991; Kauffmann, Guiderdoni & White 1994). However, what might be even more important, the fraction of gas that is able to cool early-on in small haloes will be significantly reduced.

### 3.3 Galactic substructure with scale $D_{\rm rp}$

I will now briefly speculate about the possibility that radiation-pressure support on a scale $D_{\rm rp}$ might have some observable consequences for significantly larger stellar systems. This could occur in the following way. While it is easy to see that radiation-pressure support due to ionizing radiation can not be maintained for a Hubble time, it is well possible that the collapse of a predominantly gaseous sub-system of scale $D_{\rm rp}$ is halted for several dynamical times. This will generally be sufficient to turn a considerable fraction of the gas into stars and such a sub-system might be preserved as distinct sub-structure in the later predominantly stellar system. Possible situations where this might have occurred are phases of violent star formation at the the centre of galaxies and in the the disc of spiral galaxies. Radiation pressure might therefore play a role for the formation of the distinct bulges of late-type spiral galaxies, which have masses $\sim 10^9\ M_\odot$ and radii of $\sim 1\,{\rm kpc}$ (Kent 1985, Simien & de Vauceuleurs 1986). However, it has to be admitted that the observed continuity of the dynamical properties of the bulges of late- and early-type spirals, S0 galaxies and ellipticals ( *e.g.* Bender et al. 1992) is an argument against such a scenario.

Another galactic feature with the right scale is the thick disc of our galaxy. The total surface mass density of the galactic disc is of order $10^{-2}\,{\rm g\,cm^{-3}}$ and the value for the thick disc should be a factor of a few smaller. Fig. 2 shows that for a reasonable value of $\mu_{\rm ion}/\mu_{\rm edd} \sim 10^{-2} - 10^{-3}$, the equilibrium height of the disc is then very close to $H_{\rm rp} \sim 700\,(\delta_\rho/10)$ pc which is in rather good agreement with the observed scale height of the thick disc of $\sim 1\,{\rm kpc}$ (Gilmore 1989). Radiation pressure support might render the slow collapse of the disc possible which is required in formation scenarios, where the thick disc forms early on in the evolution of the Galaxy. It might furthermore solve the problem why the thick disc has a rather high vertical velocity dispersion of $\sim 45\,{\rm km\,s^{-1}}$. This corresponds to a kinematic temperature of $\sim 2 \times 10^5$ K and is considerably higher than the characteristic $10^4$ K, where the strong drop in the cooling function occurs.

# 4 CONCLUSIONS

It was shown that absorption of ionizing radiation is an astonishingly efficient means of transferring momentum in a fully ionized self-gravitating object. There is a characteristic length scale below which momentum transfer due to ionizing radiation is the dominant force. For the simplest model, a homogeneous uniform sphere of hydrogen, it only depends on fundamental atomic physics constants

$$D_{\rm rp} = \frac{\alpha_{\rm tot}\, h\nu_{\rm ion}}{4\pi\, c\, G\, m_{\rm p}^2} \sim 110\,{\rm pc} \qquad (18)$$

In astrophysically relevant situations this fundamental length scale can be enhanced by a factor $10 - 100$ due to the clumpiness of the gas and maybe resonant line scattering in Ly-$\alpha$ and metal lines. Systems which are smaller than $D_{\rm rp}/2$ when they become ionized will be disrupted and in the case of collapsing systems, the collapse will be reverted if ionization sets in at radii smaller than $D_{\rm rp}/4$. Such a reversion of a collapse is possible for masses $\lesssim 10^9 - 10^{10}\,M_\odot$. Momentum transfer due to ionizing will therefore be a very important feedback process during the build-up of baryonic structures, especially in hierarchical cosmogonies. It is probably a more efficient means of removing baryons from small dark matter haloes in the (baryonic) mass range $10^6 - 10^9\,M_\odot$ than supernova heating. It will play an important role for the dynamical and thermal evolution of the IGM and the mass-to-light ratio of faint galaxies. Furthermore the existence of stellar substructure with scale $D_{\rm rp}$ in larger systems might be a directly observable consequence of radiation-pressure support during galaxy formation.

## ACKNOWLEDGMENTS

I would like to thank Martin Rees for helpful discussions on this topic and Leon Lucy, Peter Schneider and Matthias Steinmetz for comments on the manuscript. I acknowledge partial support by an Isaac Newton Studentship and the Gottlieb Daimler- and Karl Benz- Foundation.

# FIGURE CAPTIONS

**Fig. 1.** Relevant relations for momentum transfer by ionizing radiation in a self-gravitating inhomogeneous ($\delta_\rho = \langle n^2 \rangle / \langle n \rangle^2 = 10$) sphere of hydrogen fully ionized with the minimum necessary luminosity ($L_{\rm ion}^{\rm min}$). The thick *solid line* is the maximum radius for radiation pressure support, as discussed in section 2.1 and 2.3 ($\propto \delta_\rho$). The *long-dashed* lines are lines of constant total column density. They coincide with lines of constant neutral fraction $y$ (not dependent on $\delta_\rho$). The lower thick long-dashed line marks the limit for full ionization. Below this line the optical depth for ionizing radiation is smaller than one, even for $y \sim 1$. The upper thick long-dashed line shows the limit where the optical depth for Thompson scattering is larger than that for absorption. The *short-dashed lines* are lines of constant density and the minimum luminosity necessary for full ionization is here a constant fraction of the Eddington luminosity $\beta = L_{\rm ion}^{\rm min}/L_{\rm edd}$ ($\propto \delta_\rho$). *Dashed-dotted lines* are lines of constant mass. *Dotted lines* show the density, where baryons in an isothermal dark-matter halo are expected to become self-gravitating for different virialization redshift of the halo.

**Fig. 2.** Relevant relations for momentum transfer by ionizing radiation in a self-gravitating disc of hydrogen fully ionized with the minimum necessary surface brightness ($\mu_{\rm ion}^{\rm min}$). The thick *solid line* is the maximum disc height for radiation pressure support, as discussed in section 2.4 ($\propto \delta_\rho$). The *long-dashed lines* are lines of constant total column density. They coincide with lines of constant neutral fraction $y$ (not dependent on $\delta_\rho$). The lower thick long-dashed line marks the limit for full ionization. Below this line the optical depth for ionizing radiation is smaller than one, even for $y \sim 1$. The upper thick long-dashed line shows the limit where the optical depth for Thompson scattering is larger than that for absorption. The *short-dashed lines* are lines of constant density and the minimum surface brightness necessary for full ionization is here a constant fraction of the Eddington surface brightness $\beta = \mu_{\rm ion}^{\rm min}/\mu_{\rm edd}$ ($\propto \delta_\rho$). *Dashed-dotted lines* give the equilibrium disc height for different $\beta$ (equation (15)).

Fig. 1

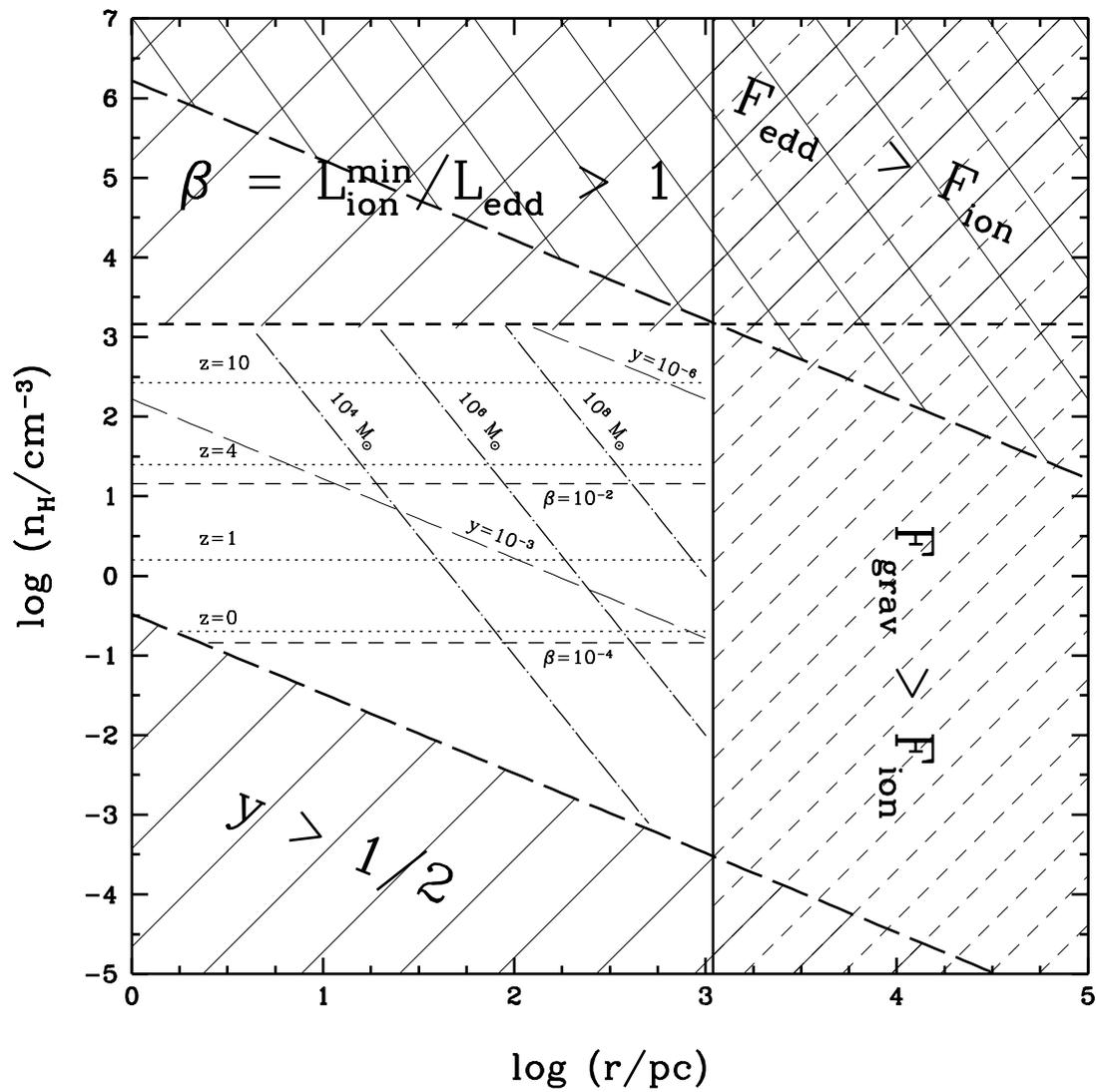

Fig. 2

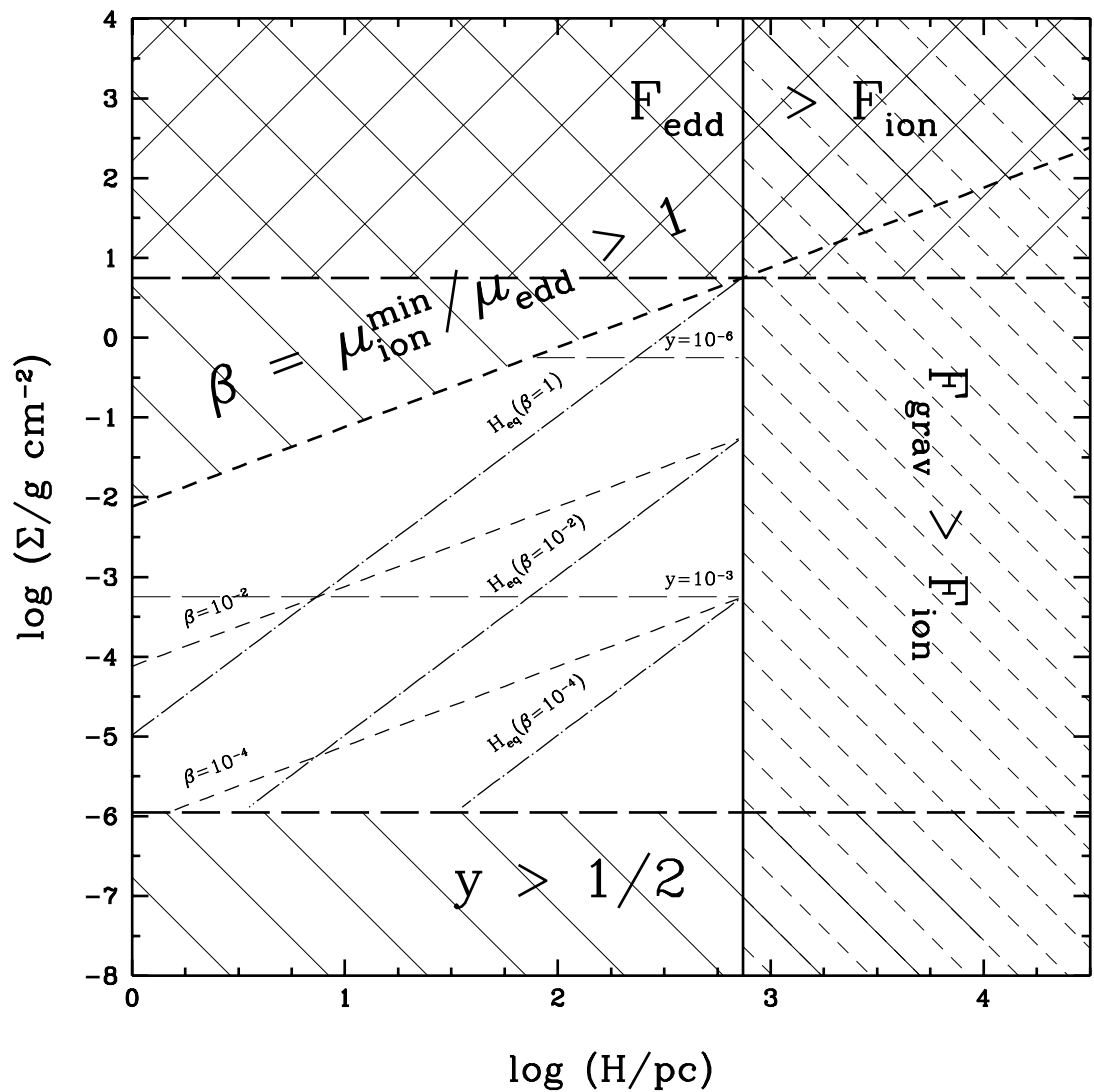